\documentclass[showpacs,twocolumn,prl,superscriptaddress]{revtex4-1}

\usepackage{graphicx}
\usepackage{dcolumn}
\usepackage{bm}
\usepackage{braket}
\usepackage[dvipdfmx]{hyperref}
\hypersetup{
colorlinks=true,
linkcolor=blue,
citecolor=blue,
urlcolor=blue,
}

\begin{document}

\preprint{preprint}
\title{Isotopic shift of atom-dimer Efimov resonances in K-Rb mixtures: \\
Critical effect of multichannel Feshbach physics}
\author{K. Kato}
\email{k\_kato@sci.osaka-cu.ac.jp}
 \affiliation{Graduate School of Science, Osaka City University, Sumiyoshi-ku, Osaka 558-8585, Japan}
 
\author{Yujun Wang}
\affiliation{Department of Physics, Kansas State University, 116 Cardwell Hall, Manhattan, KS 66506, USA}
\altaffiliation{Present address: American Physical Society, 1 Research Rd., Ridge, NY, 11961, USA}

\author{J. Kobayashi}
 \affiliation{Department of Physics, Graduate School of Science, Kyoto University, Kyoto 606-8502, Japan}

\author{P. S. Julienne}
\affiliation{Joint Quantum Institute, University of Maryland and NIST, College Park, Maryland 20742, USA}

\author{S. Inouye}
 \affiliation{Graduate School of Science, Osaka City University, Sumiyoshi-ku, Osaka 558-8585, Japan}
\date{\today}

\begin{abstract}
The multichannel Efimov physics is investigated in ultracold heteronuclear admixtures of K and Rb atoms. We observe a shift in the scattering length where the first atom-dimer resonance appears in the $^{41}$K-$^{87}$Rb system relative to the position of the previously observed atom-dimer resonance in the $^{40}$K-$^{87}$Rb system. This shift is
well explained by our calculations with a three-body model including the van der Waals interactions, and, more importantly, the multichannel spinor physics.
With only minor difference in the atomic masses of the admixtures, the shift in the atom-dimer resonance positions can be cleanly ascribed to the isolated and overlapping
Feshbach resonances in the $^{40}$K-$^{87}$Rb and $^{41}$K-$^{87}$Rb systems, respectively. Our study demonstrates the role of the multichannel Feshbach physics in determining
Efimov resonances in heteronuclear three-body systems.

\end{abstract}


\maketitle
If physical systems exhibit properties that are independent of details
of interaction, they are called universal \cite{Bra06}.
Universality has played a central role in the analysis of quantum
degenerate gases, {\em e.g.}, the effects of binary collisions were
successfully characterized by a single parameter, the $s$-wave
scattering length $a$, independent of the details of the two-body
potential.
For few-body phenomena, however, it has been well known that an
additional parameter --- {\em e.g. } three-body parameter~\cite{Bra06}
--- is necessary for a complete description of the system.
Efimov states, an infinite series of three-body bound states with
discrete scale invariance when a two-body scattering length diverges
\cite{Efi70}, provided us a unique opportunity to
investigate the properties of three-body parameter both theoretically
and experimentally. Combined experimental effort to observe
Efimov-related resonance provided us with unexpected constancy of
three-body parameters\cite{Kra06,Zac09,Wen09,Gro09,Ber11,Wil12,Dyk13}, while detailed theoretical analysis
showed
the origin of this constancy in some limiting cases\cite{Jia12,Nai14}.

Recently, a newly developed three-body spinor model that included both van der Waals interactions and multichannel Feshbach physics has succeeded in reproducing many experimentally observed Efimov features in homonuclear atomic systems \cite{Yuj14}.
This model involves additional parameters that characterize Feshbach resonances, such as the background scattering length of an open channel normalized by the van der Waals length ($r_\mathrm{bg}$) and the resonance strength of a closed channel ($s_\mathrm{res}$). 
It was impressive to see that predictions from a three-body model constructed to reproduce only two-body Feshbach physics
match almost perfectly with the experimentally observed three-body features in homonuclear systems \cite{Yuj14,Zen14}.
This achievement suggests that the necessity of including precise
few-body short-range chemical forces in studies of universal few-body phenomena --- a task far beyond our current capability --- may be removed.

Extending this universal theory to {\em hetero}nuclear systems is the next big challenge.
In addition to the mass ratio, heteronuclear systems have the extra complication of having both inter- and intra-species scattering lengths. 
Near broad Feshbach resonances ($s_{\rm res}\gg 1$), the single-channel universal van der Waals theory~\cite{Yuj12} predicted a universal dependence of the three-body parameter on
the intra-species scattering length. The predictions have been confirmed to hold in $^{6}$Li-$^{133}$Cs \cite{Shi14,Pir14,Ulmanis16} and
$^{7}$Li-$^{87}$Rb \cite{Mai15} mixtures, where the Feshbach resonances with intermediate widths have $s_{\rm res}$ even down to unity~\cite{Ulmanis16}.
For homonuclear systems, the universal prediction~\cite{Jia12} appeared to hold for $^{39}$K even near narrow Feshbach resonance with $s_{\rm res}\approx 0.1$~\cite{Roy13}. The effect of the finite Feshbach resonance width, as well as of the structure of the complicated resonance structure more commonly seen in realistic atomic systems, has therefore not been well demonstrated and understood.
Moreover, the recent experimental confirmations of the universal van der Waals theory are
for the systems with large mass ratios (two heavy and one light atoms) --- the so-called ``Efimov-favored'' systems, where the mechanism of
the universality of the three-body parameter is considered to be different
from the mechanism in the systems with small or moderate mass ratios~\cite{Yuj12} --- the ``Efimov-unfavored'' systems. An independent investigation is therefore still needed for the latter systems.

In this study, we show the role of multichannel Feshbach physics in determining the three-body parameter in the ``Efimov-unfavored'' systems of K-Rb 
admixtures. Our three-body spinor theory successfully reproduces the difference in the positions of the
Efimov-like atom-dimer resonance observed in our $^{41}$K-$^{87}$Rb experiment and the previous JILA $^{40}$K-$^{87}$Rb experiment \cite{Blo13}. 
As the Feshbach resonances
involved in the isotopic systems include both isolated and overlapping resonances, our study also demonstrates that Efimov resonance features
in atomic systems can be fully described by three-body van der Waals models that reproduce the relevant two-body Feshbach spectra. 

In previous experiments, the three-body parameter was mostly determined by the positions of the Efimov resonances in three-body recombination.
This approach is feasible in studies of the ``Efimov-favored'' heteronuclear systems, thanks to the increased universal scaling constant $s_0$ and therefore,
the significantly decreased Efimov scaling cycle $e^{\pi/s_0}$ ~\cite{Bra06} for the ground Efimov resonance to be observed. For an ``Efimov-unfavored'' system such as
Rb-Rb-K, the size of scattering length needed for seeing an Efimov resonance in three-body recombination is too large to be realized experimentally.
We therefore measure the positions of the Efimov-like atom-dimer resonances instead, which could be observed at significantly lower scattering length~\cite{Hel10}.
For a comprehensive analysis of the heteronuclear Efimov resonance,
we also investigate the three-body loss in $^{41}$K-$^{87}$Rb admixtures on the negative side of the Feshbach resonance. As will be shown in our later discussion, the absence of a
resonance in three-body recombination in our measurements is consistent with the universal predictions~\cite{Yuj12}. 

There are several experimental groups working on different isotopes for the KRb mixture \cite{Bar09,Blo13,Min14,Wac16}. We compare our results with those obtained by the JILA group for $^{40}$K-$^{87}$Rb mixture \cite{Blo13,Min14}.
Two KRb systems are nearly identical in the single channel theory: the van der Waals length ($r_\mathrm{vdW}$) are 71.9 $a_0$ and 72.2 $a_0$ \cite{sim08}, and the scaling parameters $s_0$ for two Rb atoms and one K atom are 0.6444 and 0.6536, for $^{41}$K-$^{87}$Rb mixture and $^{40}$K-$^{87}$Rb mixture, respectively\cite{Hel10}.
However, they are different in the multi-channel theory: the mixture of $^{41}$K atoms and $^{87}$Rb atoms both in the lowest hyperfine state have one broad resonance at 39.4 G and one intermediate resonance at 78.82 G.
Special treatment is needed for describing these overlapping Feshbach resonances \cite{SupMat}.
The mixture of $^{40}$K atoms and $^{87}$Rb atoms both in the lowest hyperfine state have one intermediate resonance at 546.9 G.
This resonance is well isolated from other Feshbach resonances and thus it can be simply described by the combination of $s_\mathrm{res}$ ($=1.96$) and $r_\mathrm{bg}$ ($=-2.75$)\cite{Chi10}. 
The purpose of comparing these two KRb systems is twofold. First, if these two systems show identical Efimov structures, it is very likely that a single-channel theory can adequately explain the three-body physics in heteronuclear systems; 
conversely, if the results do not match, it should be investigated whether the difference can be reproduced by using the multichannel van der Waals theory introduced in the homonuclear case.
 
The details of our experimental setup can be found in Ref.\ \cite{Kis09,Kato_preparation}.
In summary, we prepared a dual-species Bose-Einstein condensate (dual-BEC) comprising $^{41}$K atoms and $^{87}$Rb atoms in a crossed optical dipole trap.
Both atoms are in the $\left| F,m_F\right>=\left|1,1\right>$ state, where $F$ corresponds to the atomic angular momentum and $m_F$ is its projection. 
The typical number of each of the $^{87}$Rb and $^{41}$K atoms in a dual-BEC is $\sim 4 \times 10^4$. 

When it comes to the study of inelastic atom-dimer collisions, it is necessary for dimers to be produced efficiently. This is especially true for a Bose-Bose mixture, in which a large contribution from
atom-dimer and dimer-dimer inelastic collisions limits the efficiency with which the dimers in traps are produced.
We resolve this problem by using a three-dimensional optical lattice. Some preparatory steps are needed before the dual-BEC can be loaded onto the optical lattice potential. First, we compensate for the differential gravitational sag between the $^{41}$K and $^{87}$Rb atoms by introducing an additional laser beam whose wavelength is 809 nm \cite{S_Ospelkaus_PhD}. Second, we decompress the BEC by decreasing the trapping frequencies in the horizontal directions. This is necessary for increasing the number of lattice sites that have exactly one K and one Rb atom when they become a dual Mott insulator phase.
Typical trap frequencies for K and Rb are $(f_x,f_y,f_z)=(13, 92, 38)$ Hz and $(10, 92, 38)$ Hz, respectively, where the $y$ axis is the axis of gravity.
Finaly, we set the magnetic field $B=85$ G, where interspecies scattering length $a = -20 \,a_0$.
At this magnetic field, the dual BEC is miscible because the interspecies scattering length is much smaller than the intra-species scattering length ($a_\mathrm{KK}=63.5(6)\,a_0$ \cite{Fal08}, $a_\mathrm{RbRb}=100.4(1)\,a_0$ \cite{Kem02}). 

As we raised the optical lattice potential, the dual BEC was transformed into a dual Mott insulator. Then magnetic field was swept across the Feshbach resonance at 78.82 G
and the atoms were adiabatically associated into molecules.
For measuring the atom(Rb)-dimer(KRb) loss coefficient, we selectively removed the K atoms by applying a radio-frequency spin-flip followed by a radiative force from a laser beam tuned to the closed transition for K atoms \footnote{Although our lattice potentials were sufficiently high to prevent Rb atoms and KRb molecules from tunneling through to the neighboring sites, the tunneling rate
of K atoms was not negligible.  By removing K atoms, the lifetime of the molecules in a lattice reached 210(15) ms, which was adequate for the magnetic field for loss measurements to be set.}.
Before imaging, the atoms and molecules were spatially separated in the horizontal direction via application of a magnetic field gradient during a time-of-flight.
Furthermore, molecules were dissociated into atoms by sweeping the magnetic field across Feshbach resonance.
The typical numbers of Rb atoms and KRb molecules are $\sim 1.1 \times 10^4$ and $\sim 3 \times 10^3$, respectively.

\begin{figure}
\includegraphics[clip,width=8.0cm]{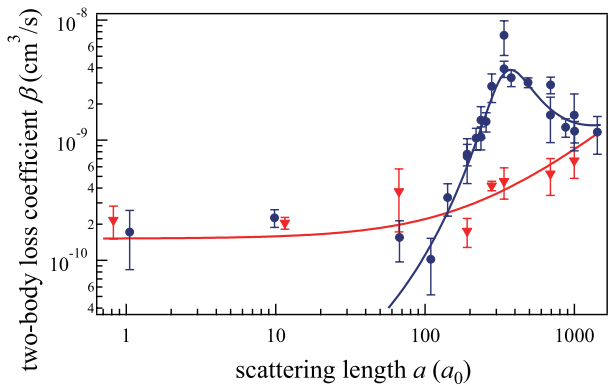}
\caption{Atom-dimer and dimer-dimer
loss coefficients observed in an ultracold $^{41}$K-$^{87}$Rb system. While the dimer-dimer (KRb-KRb) loss coefficient (red triangles) does not show any prominent feature, the atom-dimer (Rb-KRb) loss coefficient (blue circles) shows a resonant feature. Lines of best fit are illustrated by solid lines. The fit for the dimer-dimer loss coefficient assumes a linear dependence on $a$, while the fit for the atom-dimer loss assumes equation (\ref{eq:beta_ad}). The typical initial densities of the atoms and dimers are $n_\mathrm{Rb}=1.3 \times 10^{11}$\,$\mathrm{cm^{-3}}$and $n_\mathrm{KRb}=0.7 \times 10^{11}$\,$\mathrm{cm^{-3}}$, respectively.
The typical temperature is 60 nK.
}\label{fig:atom_dimer} 
\end{figure}

The atom-dimer loss coefficient $\beta_\mathrm{ad}$ was determined by placing the atom-dimer mixture into a dipole trap and measuring the number of dimers and atoms after a variable holding time $t$.  The rate equation for the number of dimers $N_\mathrm{KRb}$ can be expressed as follows:
\begin{equation}\begin{array}{lcl}
\dot{N}_\mathrm{KRb}(t)&=&-\beta_\mathrm{ad} \int n_\mathrm{Rb}(\mbox{\boldmath $r$},t)n_\mathrm{KRb}(\mbox{\boldmath $r$},t) 
d^3\mbox{\boldmath $r$}\\
&&-2\beta_\mathrm{dd} \int n_\mathrm{KRb}(\mbox{\boldmath $r$},t)^2 
d^3\mbox{\boldmath $r$} .
\end{array}\label{eq:dimer}
\end{equation}
In this equation, $N_\mathrm{Rb}(t)$ and $N_\mathrm{KRb}(t)$ are the number of $\mathrm{Rb}$ atoms and $\mathrm{KRb}$
dimers, respectively; $n_\mathrm{Rb}(\mbox{\boldmath $r$},t)$ and $n_\mathrm{KRb}(\mbox{\boldmath $r$},t)$ are the density of the $\mathrm{Rb}$ atoms and $\mathrm{KRb}$
dimers, respectively; and $\beta_{ad}$ and $\beta_{dd}$ are the
loss coefficient for the atom-dimer and dimer-dimer collisions, respectively.
Assuming a thermal distribution for the atoms and
dimers in the dipole trap, the right-hand side of equation (\ref{eq:dimer}) can be calculated using the number and temperature of atoms and
dimers from the time-of-flight images. Both $\beta_\mathrm{ad}$ and $\beta_\mathrm{dd}$ can be evaluated by comparing the experimental data from different initial conditions \footnote{By preparing a pure molecular sample, $\beta_\mathrm{dd}$ can be measured exclusively. A molecular sample can be prepared
by tuning the magnetic field to $\sim 77$ G, and then applying a magnetic field gradient.
Since the magnetic moment of the molecule is zero at 77 G,
the K and Rb atoms can be removed.
From the data with finite $N_\mathrm{Rb}(t)$, 
$\beta_\mathrm{ad}$ can be determined by subtracting the contribution from $\beta_\mathrm{dd}$.}.
\begin{table}
\caption{
Experimental results of the atom-dimer resonances for Rb+$^{41}$KRb  and Rb+$^{40}$KRb collisions.
These values are determined by fitting of equation (\ref{eq:beta_ad}). Subscripts of ``sys'' and ``fit'' denote systematic error and fitting error, respectively.}
\label{tab:fit_results}
\begin{tabular}{c|ccc}
\hline \hline
 & $a_\ast$ & $\eta_\ast$ & $C_\beta$ \\
\hline
$^{41}$K-$^{87}$Rb & 348(8)$_\mathrm{fit}$(41)$_\mathrm{sys}$ & 0.24(2)$_\mathrm{fit}$ & 15.5(7)$_\mathrm{fit}$ \\
$^{40}$K-$^{87}$Rb\cite{Blo13} & 230(10)$_\mathrm{fit}$(30)$_\mathrm{sys}$ & 0.26(3)$_\mathrm{fit}$ & 3.2(2)$_\mathrm{fit}$ \\
\hline \hline
\end{tabular}
\end{table}
Figure \ref{fig:atom_dimer} shows the measured
atom-dimer and dimer-dimer 
loss coefficient, $\beta$.
The magnetic field was converted into the scattering length by using the a(B) from our multichannel two-body calculation.
The calculation uses
the atomic potentials in Refs.~\cite{Pas07,Kle07} and is calibrated carefully to give the correct positions of the Feshbach resonances. 
The dimer-dimer loss coefficient $\beta_{dd}(a)$ does not show any prominent features 
\footnote{By interpolating $\beta_\mathrm{dd}(a)$ with a linear function, we obtained $\beta_\mathrm{dd}(a)=b_1 (a/a_0) +b_2$, where
$b_1=7(2)\times 10^{-13}$ cm$^3$/s and $b_2=1.5(6)\times 10^{-10}$ cm$^3$/s.}.
The resonant feature was clearly observed in the atom-dimer loss coefficient $\beta_{ad}(a)$, and the overall shape of the resonance was quite similar to that of the $^{87}$Rb-$^{40}$K$^{87}$Rb mixture. The peak position, however, was different.
We can quantify the difference in the peak positions by fitting the curves with the results obtained from the effective field theory, which includes three fitting parameters ($a_{\ast}$, $\eta_{\ast}$, $C_{\beta}$) \footnote{This form is expected to be valid when $a>2r_\mathrm{vdW}$ \cite{Hel10}.}.
\begin{equation}
\beta_{ad}(a)=C_\beta \frac{\mathrm{sinh} (2\eta_\ast)}{\mathrm{sin} ^2[s_0 \mathrm{ln} (a/a_\ast)]+\mathrm{sinh}^2(\eta_\ast )}\frac{\hbar a}{m_1}\label{eq:beta_ad}
\end{equation}
In equation (\ref{eq:beta_ad}), $a_\ast$ represents the resonance position, $\eta_\ast$ is the resonance width, and $C_\beta$ is the overall magnitude of the loss.
Note that $m_1$ is the mass of the K atom in this case, and $s_0$ is the scaling parameter.
We can fit $\beta_\mathrm{ad}$ using equation (\ref{eq:beta_ad}) and compare the results with those obtained for the $^{40}$K-$^{87}$Rb system. The results of both fits are summarized in Table \ref{tab:fit_results}. 

An isotopic comparison showed that $\eta_{\ast}$ matches within the error bars, while $a_{\ast}$ and $C_{\beta}$ are different between the two isotopes.
The difference in $C_\beta$ 
can be attributed to the systematic uncertainty in density calibration, whereas
the difference in $a_{\ast}$ \footnote{The systematic error of $a_{\ast}$ comprises a 
fluctuation in the magnetic field and the uncertainty of $a$-to-B conversion.}
signifies the difference in the position of the peak. Thus, it is worth asking whether the difference can be attributed to the difference of the properties of the Feshbach resonances.

To better understand why there was an isotopic difference, we also checked the three-body recombination rate.
Recent studies on the three-body recombination coefficient of the K-Rb systems \cite{Blo13,Min14,Wac16} showed that there is no Efimov-related
resonance in the region of $200\, a_0 < |a| < 3000\, a_0$.
Furthermore, the single-channel universal theory on the heteronuclear Efimov resonance for broad resonance
predicts that there is no resonance in the region of $|a| < 2800\,a_0$ \cite{Yuj12}.

\begin{figure}
\includegraphics[clip,width=8.0cm]{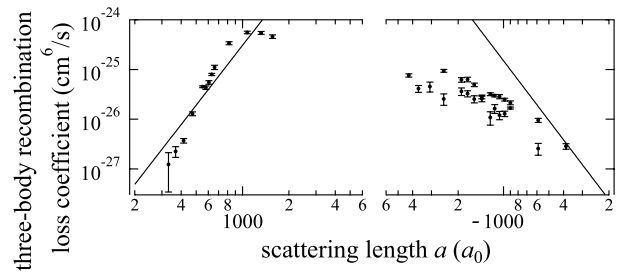}

\caption{The three-body recombination loss coefficient of the K-Rb-Rb collision in the vicinity of the heteronuclear Feshbach resonance (circles). No Efimov-like feature was observed. The solid line shows an $a^4$ dependence, and the amplitude factor is determined by a fitting on the range of the positive scattering length $200\, a_0 <a< 2000\, a_0$. On the negative side, the amplitude factor is half of its value on the positive side \cite{Hel10}. 
The $a^4$ dependence on the negative side was observed for $|a| < 2000\, a_0$, and it saturates because of a finite density. The typical density and temperature of the cloud are approximately $1\times 10^{13}$ cm$^{-3}$ and 400 nK, respectively. This corresponds to 
a thermal de Broglie length of $\lambda_\mathrm{dB}=7000 \, a_0$ and
a reciprocal wavenumber $k^{-1}=1/(6\pi n)^{1/3}=3000 \, a_0$.
}\label{fig:TBC}
\end{figure}
Experimental details on how the three-body loss coefficient was measured will be presented elsewhere \cite{Kato_preparation}. Measuring the three-body loss coefficient for a heteronuclear system of bosonic atoms is problematic because we have to distinguish between competing processes. In the case of the three-body loss for the $^{41}$K-$^{87}$Rb mixture in the vicinity of the $^{41}$K-$^{87}$Rb Feshbach resonance, there are two major contributions: K-K-Rb and K-Rb-Rb \footnote{There are contributions from homonuclear three-body loss as well. They are expected to be of the order of $10^{-29}$ cm$^6$/s 
\cite{Kis09,Sod99} and fairly small compared with the contributions from the heteronuclear loss.}.  Therefore, increasing the signal-to-noise ratio of data is mandatory. In our experiment, the main source of noise in the data analysis originated from fluctuations in the  initial number of atoms. 
We eliminated these fluctuations by taking multiple images of the same cloud using phase-contrast imaging. Additionally, we enhanced the three-body loss by increasing the atomic density \footnote{The typical trap frequencies for K and Rb are $(f_x, f_y, f_z)=(120,330,310)$ Hz and $(90,320,310)$ Hz, respectively.
The typical time scales for the loss measurement are of the order of several tens of milliseconds.}.
The observed three-body loss coefficient (Fig.\ref{fig:TBC}) did not show any Efimov-like structures \cite{Wac16}. 

\begin{figure}
\includegraphics[clip, width=8.0cm]{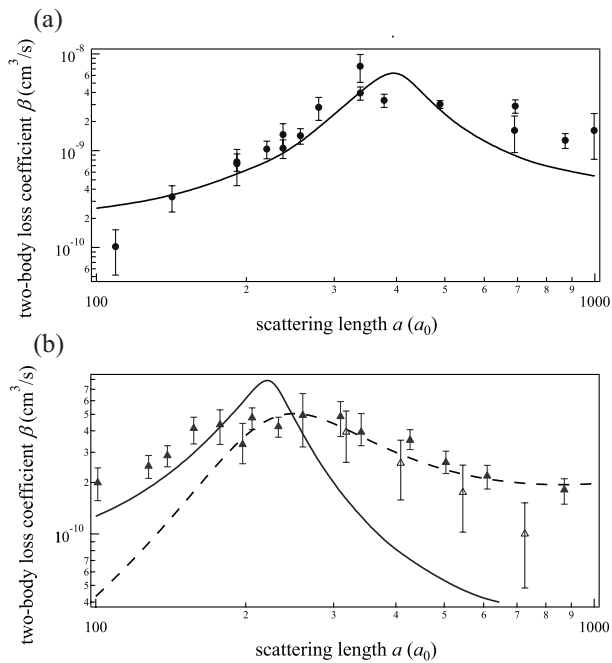}
\caption{
Comparisons of numerically calculated and experimentally measured loss
coefficients for (a) Rb+$^{41}$KRb  and (b) Rb+$^{40}$KRb (experimental data obtained from Ref. \cite{Blo13}) collisions.
In both graphs, results from numerical calculations (shown in solid
lines) show reasonable agreement with experimental results (shown in circles and triangles). The
result from the fit using the effective field theory (dotted line) is
also shown in (b). The temperature of each measurement was $\sim$60 nK
(solid circles), $\sim$150 nK (open triangles), and $\sim$300 nK
(solid triangles), respectively. The theoretical results (solid lines)
are multiplied by 2 and 5, 
which are thermally averaged 
at 70 nK and 300 nK
for Rb+$^{41}$KRb and Rb+$^{40}$KRb loss coefficients,
respectively.
The peak positions of theoretical curves are 395 $a_0$ and 222 $a_0$ for Rb+$^{41}$KRb and Rb+$^{40}$KRb loss coefficients, respectively.
The resonance position in Rb+$^{41}$KRb loss coefficients shown in
(a) are insensitive to the temperature change in the range of
temperature we study, whereas the calculated coefficients shown at 70 nK give
the best agreement with the line shape of the measured loss coefficients at
lower scattering lengths. 
}
\label{fig:from_Yujun} 
\end{figure}

The significant shift in the positions of the atom-dimer resonances in the two isotopic K-Rb admixtures clearly cannot be explained by the 
small differences in the van der Waals lengths or the universal scaling constants. In fact, the single-channel, universal van der Waals three-body theory 
fails here by a large margin with an incorrectly predicted atom-dimer resonance near $a=$100 $a_0$. The failure of the universal theory raises the important question 
of whether the general universality, \textit{i.e.}, the independence of Efimov physics on short-range chemical forces,
is still valid. 

Our approach to address the question above
is to perform three-body calculations with a spinor model that reproduces the relevant two-body Feshbach spectra in each of the isotopic systems. 
Such a model is relatively easy to build in the Rb-Rb-K system, because the multichannel physics is only important for the K-Rb pairs, whereas a single-channel 
description is a good approximation for the Rb-Rb interaction in the whole range of magnetic field of our current interest. Our theory therefore allows the K atom to carry 
the (pseudo)spin degrees of freedom and treats Rb atoms as spinless. Specifically, the total three-body wave function $\Psi$ is expanded as 
$\Psi=\sum_\alpha\psi_\alpha |\alpha\rangle$, where $|\alpha\rangle$ is the (pseudo)spin state of the K atom.

With the spinor model, we solve the three-body Schr\"odinger equation in the form of
\begin{equation}
(T+V_{\rm RbRb})\psi_{\alpha}+\sum_\beta (V_{\rm KRb,1}^{(\alpha\beta)}+V_{\rm KRb,2}^{(\alpha\beta)})\psi_\beta=(E-\epsilon_\alpha)\psi_\alpha,
\label{Eq_3b}
\end{equation}
where $T$ is the three-body kinetic energy operator, $V_{\rm RbRb}$, $V_{\rm KRb,1}^{(\alpha\beta)}$, and $V_{\rm KRb,2}^{(\alpha\beta)}$ the single- and multi-channel 
two-body potentials of the three pairs of the atoms, and $\epsilon_\alpha$ the single-atom energy level of the K atom.
The proper magnetic-field dependence of the three-body Hamiltonian is built in the single-atom energy as $\epsilon_\alpha=\mu_\alpha B+u_\alpha$,
where the magnetic moment $\mu_\alpha$ and the zero-field energy $u_\alpha$ are chosen to
mimic the realisitc magnectic moments and the hyperfine splittings~\cite{SupMat}.
We solve Eq.~(\ref{Eq_3b}) and 
calculate the atom-dimer loss coefficients with essentially the same potential models and numerical techniques used in Ref.~\cite{Yuj14}. 

In Figure~\ref{fig:from_Yujun} we show our numerically calculated atom-dimer loss rates compared with the data from our and JILA's experiments. 
To properly reproduce the isolated and overlapping characters of the Feshbach resonances in $^{40}$K-Rb and $^{41}$K-Rb pairs, 
two-spin-state and three-spin-state model interactions are used for $V_{\rm KRb}^{(\alpha\beta)}$ in the $^{40}$K-Rb-Rb and $^{41}$K-Rb-Rb calculations, respectively. 
Without fitting parameters, the calculated atom-dimer resonance positions agree well with both of the experimentally observed positions, and consequently, 
reproduce the atom-dimer resonance shift in the isotopic systems. 

For the $^{41}$K-Rb-Rb system, we point out that in order to correctly predict the atom-dimer resonance position,
it is necessary for the three-body model to reproduce both the background 
(39 G) and overlapping (79G)
Feshbach resonances. A model that
reproduces only the local properties of the overlapping resonance (\textit{i.e.}, the local $r_{bg}$ and $s_{\rm res}$) does not give the atom-dimer
resonance position correctly. Another observation is that regardless of the number of spin states,
the calculated loss rates in the two isotopic systems have similar magnitude when the scattering length is low.  
This suggests that the observed shift of the atom-dimer resonance position --- going beyond the single-channel van der Waals theory --- is the manifestation of the difference in the underlying two-body Feshbach physics.
The short-range chemical forces are clearly not involved.

In summary, we measured the heteronuclear atom-dimer loss coefficients of $^{87}$Rb atoms and $^{41}$K$^{87}$Rb Feshbach molecules at ultracold temperatures. The observed loss coefficient showed an Efimov-related resonance at
$a_\ast = 348(8)_\mathrm{fit}(41)_\mathrm{sys}$\,$ a_0$
, which shifted from previous measurements for different isotopes of potassium.  To explain this shift, we modeled the system using a 
three-body spinor theory that reproduced the properties of Feshbach resonances.
This theory was successful in reproducing the experimental results of the atom-dimer resonance for both isotopes.
These results show the important role of the multichannel Feshbach physics in shifting the positions of the three-body Efimov resonances, and 
demonstrates the independence of these three-body resonances on the short-range chemical forces in the heteronuclear atomic systems even near 
relatively narrow Feshbach resonances.

The authors would like to thank Shimpei Endo and Pascal Naidon for their valuable suggestions.
This work was supported by JSPS KAKENHI Grant-in-Aid for Scientific Research(B), Grant Number JP23340117.

%

\widetext
\clearpage
\begin{center}
\textbf{\large Supplemental Material for ``Isotopic shift of atom-dimer Efimov resonances in K-Rb mixtures:
Critical effect of multichannel Feshbach physics
''}
\end{center}
\setcounter{equation}{0}
\setcounter{figure}{0}
\setcounter{table}{0}
\setcounter{page}{1}
\makeatletter
\renewcommand{\theequation}{S\arabic{equation}}
\renewcommand{\thefigure}{S\arabic{figure}}
\renewcommand{\bibnumfmt}[1]{[S#1]}
\renewcommand{\citenumfont}[1]{S#1}
%
%
%
%
\begin{figure}[h]
\includegraphics[clip,width=16.0cm]{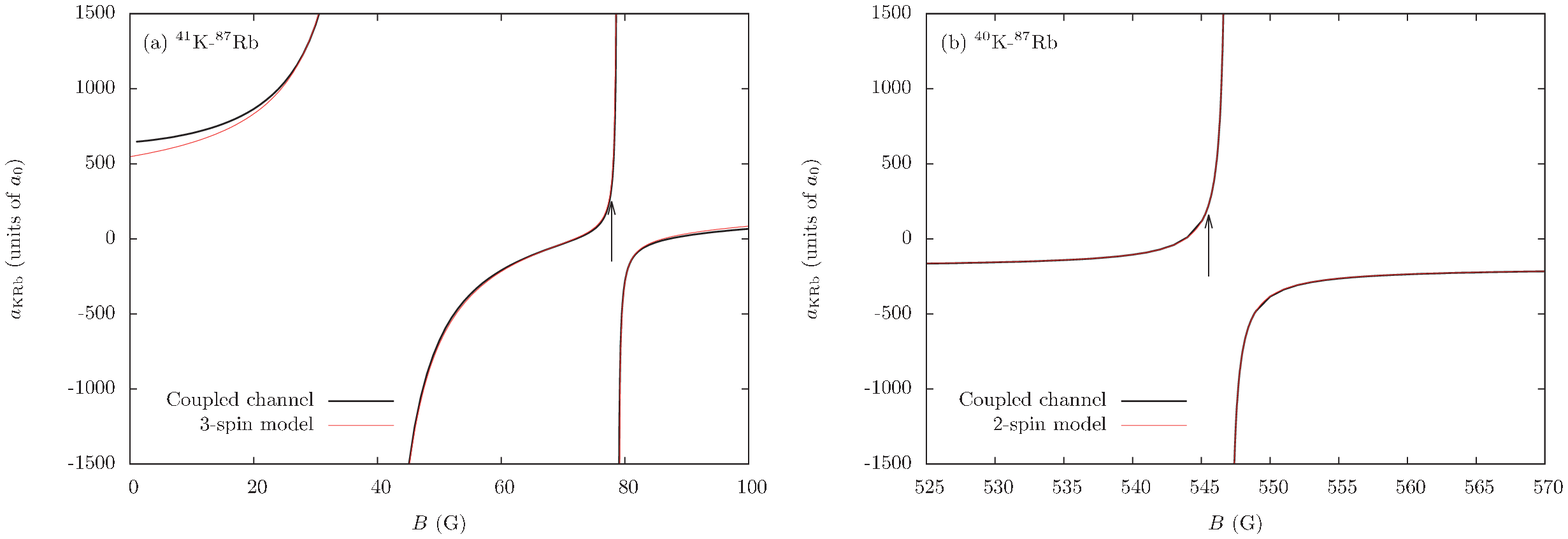}
\caption{Scattering length from coupled-channel two-body calculations and calculations with our spinor models, (a) near the 
$^{41}$K-$^{87}$Rb 39 G and 79 G overlapping Feshbach resonances, and (b) near the $^{40}$K-$^{87}$Rb 547 G $s$-wave Feshbach resonnace.  
The arrows indicate the positions of the atom-dimer resonances observed in the experiments.}
\label{fig:SM_3B}
\end{figure}
\section{Atomic interaction models used in three-body calculations}

To properly represent the multichannel physics of the Feshbach resonances between K and Rb atoms in our three-body calculations, we allow the K atom to carry 
the spin degrees of freedom with a spin state $|\alpha\rangle$. The potential between K and Rb atoms $V_{\rm KRb}^{\alpha\beta}(r)$, where $r$ is the distance 
between the K and Rb atoms, is constructed to reproduce the 
long-range ($r\gtrsim 20$ bohr) behavior of the two-body wave function near the relevant Feshbach resonances. 

We use Lennard-Jones 6-12 potentials with the same $C_6$~\cite{Pas07Sup,Kle07Sup} for the diagonal potentials $V_{\rm KRb}^{\alpha \alpha}(r)$. 
The open-channel potentials (associated with the lowest zero-field energy $u_\alpha$) 
have short-range cutoffs such that the single-channel scattering lengths reproduce the background scattering lengths 
near the Feshbach resonances investigated in our current study. For the $^{40}$K-$^{87}$Rb interaction, this open-channel scattering length is $-189$ bohr, 
which corresponds to the background scattering length of the isolated $s$-wave Feshbach resonance near $547$ G. 
For the $^{41}$K-$^{87}$Rb interaction, the open-channel scattering length 
is $284$ bohr, which is the global background scattering length of the overlapping Feshbach resonances near 39 and 79 G.

The closed-channel potential(s) is constructed in the way such that its bound-state wave function reproduces the main characters of the closed-channel wave function 
found in the coupled-channel two-body calculations. In both $^{40}$K-$^{87}$Rb and $^{41}$K-$^{87}$Rb systems, 
the closed-channel component 
of the low-energy two-body scattering wave function is found to have a strong character of the triplet $-2$ vibrational level of the corresponding system. 
Based on this observation, we adjust the short-range cutoffs of the closed-channel potentials of $^{40}$K-$^{87}$Rb and $^{41}$K-$^{87}$Rb 
systems such that the potentials have triplet scattering lengths of $-214$ and $164$ bohr, respectively, and use the 
the $-2$ vibrational levels of the closed channels to give rise to the Feshbach resonances in our spinor models. 

The single-atom energies $u_\alpha$ are chosen for each spin channel to 
be on the order of the hyperfine splittings between the relevant K-Rb scattering thresholds, and more importantly, 
to accurately position the Feshbach resonances in our spinor models. The magnetic moments $\mu_\alpha$ are set to reproduce the magnetic moment 
differences between the open and closed channels found in coupled-channel two-body calculations. 

In Figure~\ref{fig:SM_3B}, we show the scattering length from the coupled-channel calculations and those from our two- and three-spin models used in our 
three-body calculations. The coupled-channel calculations include only the $s$-wave basis. For the $^{40}$K-$^{87}$Rb 
system, although 
there is a $d$-wave Feshbach resonance located at about 0.8 G higher than the $s$-wave resonance, 
this $d$-wave resonance is very narrow~\cite{Bra13} and produces very small perturbation to 
the $s$-wave Feshbach resonance. Moreover, the atom-dimer resonance was observed on a branch of the $s$-wave Feshbach resonance different 
from the branch where the $d$-wave resonance locates, therefore the effect of the $d$-wave resonance can be safely neglected in our study of the 
$^{40}$K-$^{87}$Rb-$^{87}$Rb system.
No significant higher partial-wave resonances are found in the $^{41}$K-$^{87}$Rb 
system in the range of magnetic field in our current study.

\end{document}